%% file: ms.tex
\documentclass[runningheads]{llncs}
\usepackage{graphicx}
\usepackage{soul}
\usepackage{todonotes}
\usepackage{booktabs}
\usepackage[utf8x]{inputenc}
\usepackage{tabularx}
\usepackage{paralist}
\usepackage[normalem]{ulem}
\usepackage{listings}
\usepackage{hyperref}
\newtheorem{ex}{Example}

\newcommand{\figref}[1]{Fig.\,\ref{#1}}
\newcommand{\tabref}[1]{Table\,\ref{#1}}

\newcommand{\exref}[1]{Example\,\ref{#1}}

\begin{document}
\title{Investigating Software Usage in the Social Sciences: A Knowledge Graph Approach}

\titlerunning{Investigating Software Usage in the Social Sciences with SoftwareKG}

\author{David Schindler\inst{1}\orcidID{0000-0003-4203-8851} \and
Benjamin Zapilko\inst{2}\orcidID{0000-0001-9495-040X} \and
Frank Krüger\inst{1}\orcidID{0000-0002-7925-3363}}
\authorrunning{D. Schindler et al.}

\institute{Institute of Communications Engineering, University of Rostock, Rostock, Germany
\email{first.last@uni-rostock.de}\\
\url{https://www.int.uni-rostock.de/} \and
GESIS - Leibniz Institute for the Social Sciences, Cologne, Germany\\
\email{Benjamin.Zapilko@gesis.org}\\\url{https://www.gesis.org/}}
\maketitle             
\begin{abstract}
Knowledge about the software used in scientific investigations is necessary for different reasons, including provenance of the results, measuring software impact to attribute developers, and bibliometric software citation analysis in general.
Additionally, providing information about whether and how the software and the source code are available allows an assessment about the state and role of open source software in science in general.
While such analyses can be done manually, large scale analyses require the application of automated methods of information extraction and linking.
In this paper, we present SoftwareKG --- a knowledge graph that contains information about software mentions from more than 51,000 scientific articles from the social sciences.
A silver standard corpus, created by a distant and weak supervision approach, and a gold standard corpus, created by manual annotation, were used to train an LSTM based neural network to identify software mentions in scientific articles.
The model achieves a recognition rate of .82 F-score in exact matches.
As a result, we identified more than 133,000 software mentions.
For entity disambiguation, we used the public domain knowledge base DBpedia.
Furthermore, we linked the entities of the knowledge graph to other knowledge bases such as the Microsoft Academic Knowledge Graph, the Software Ontology, and Wikidata.
Finally, we illustrate, how SoftwareKG can be used to assess the role of software in the social sciences.


\keywords{software in science  \and scientific articles \and information extraction \and knowledge graph.}
\end{abstract}
\section{Introduction}
Software is used during the entire research life-cycle and thus has significant influence on the research and its results.
Knowledge about the software that was used during a scientific investigation is of interest for various reasons~\cite{Krueger2019}, for instance to track the impact of software to attribute its developers, to analyze citation patterns, or to assess provenance information with respect to the research workflow.
This is of particular interest, as software might contain issues that can affect the scientific results, as for instance reported by \cite{eklund2016cluster,Ziemann2016,Zeeberg2004}.
Moreover, research often relies on closed source software which is often not fully validated~\cite{Russo2016}, eventually creating uncertainty about the reliability of the results.

Recently, software citation standards~\cite{Smith2016} have gained increasing interest in the scientific community but are not consistently used, which hampers the identification of such information on a large scale.
Moreover, researchers are surprisingly creative when it comes to spelling variations of the actual software name (see \tabref{tab:spss_variations}).
Named entity recognition (NER) provides a convenient method to identify entities in textual documents and could thus be employed to extract software mentions from scientific articles. 
The objective is to identify all software with an assigned name while ignoring unspecific statements such as `custom script' or `custom code'.
We employed an LSTM based approach to NER, which was trained using transfer learning based on distant and weak supervision and a small corpus of manually annotated articles~\cite{Schindler_softwarekg_2020}.
The resulting information was then structured, interlinked, and represented in a knowledge graph which enables structured queries about software mentions in and across scientific publications.
By exploiting this information as a knowledge graph, we follow W3C recommendations and best practices~\cite{heath2011linked}.

In this paper, we present SoftwareKG, a knowledge graph that links 51,165 scientific articles from the social sciences to software that was mentioned within those articles.
SoftwareKG is further curated with additional information about the software, such as the availability of the software, its source and links to other public domain knowledge graphs.
Using this information and exploiting additional information via links to other knowledge graphs, SoftwareKG provides the means to assess the current state of software usage, free and open source software in particular, in the social sciences.
Links to other knowledge bases play an important role since additional information about software and scientific articles can be accessed which is not available directly from the article.
All software and data associated with SoftwareKG is publicly available~\cite{Schindler_softwarekg_2020}.

The remainder of this paper is structured as follows: 
First we describe how the information from articles was extracted, curated and structured. 
Afterwards, we provide a brief description of SoftwareKG including entity and relation statistics. 
We then discuss potential error sources and illustrate how SoftwareKG could be employed for the analysis of software usage in the social sciences.
Finally, we discuss related work, summarize, conclude and lay out potential further work.  

\section{Document Selection and Corpus Generation}

\subsection{Gold Standard Corpus Generation}\label{sec:gsc}
The gold standard corpus (GSC) was created by randomly selecting 500 articles from PLoS\footnote{\url{https://www.plos.org/}} using the keyword ``Social Science''.
All articles were scanned for Methods \& Materials (M\&M) sections, based on the assumption that those sections contain most of the software usage statements~\cite{duck2013bionerds}.
From the initial set of 500 articles, M\&M sections were found in and extracted from 480, which then served as a base for the GSC.
The remaining articles did not contain a M\&M or similar section and were thus omitted.
Ground truth annotation was performed by seven annotators using the BRAT v1.3~\cite{stenetorp2012brat} web based annotation tool.
Annotators were instructed to label all software usage statements, excluding any version or company information (see Example~\ref{ex3}).
Inter-rater reliability was assessed on 10\% of all sentences and showed almost perfect agreement~\cite{landis1977measurement} (Cohen's $\kappa{=}.82$).
The GSC was then split into training, development and test sets with relative amounts of 60\%, 20\%, and 20\%, respectively.
The training set was extended by 807 sentences with software names to increase the amount of positive samples.
The set of positive samples was retrieved by selecting sentences that contain at least one of the 10 most common software names of a previous analysis~\cite{duck2013bionerds}.
An overview of the resulting GSC is provided in \tabref{tab:gsc}. 
The GSC is publicly available~\cite{Schindler_softwarekg_2020}.
\begin{table}[tb]
    \centering
    \caption{Overview of the GSC.}
    \label{tab:gsc}
    \begin{tabularx}{.8\textwidth}{Xp{2cm}p{2cm}Xp{1cm}}
    \toprule
    \multicolumn{2}{c}{GSC Statistics} & &\multicolumn{2}{c}{Most frequent software}\\
    \midrule
    \# Sentences & 31,915 & &R & 77\\
    \# Annotations & 1380 & &SPSS & 60\\
    \# Distinct & 599 & &SAS & 44\\
    Train & 847 (+1005) & &Stata & 41\\
    Devel & 276 & &MATLAB & 38 \\
    Test & 257 & &Matlab & 28 \\ 
    \bottomrule
  \end{tabularx}  
\end{table}

\subsection{Silver Standard Corpus Generation}
Named entity extraction methods, in particular those relying on neural networks, require large amounts of training data to achieve reasonable recognition results.
Annotated training data, however, is often unavailable and expensive to produce.
The application of silver standard corpora~\cite{REBHOLZ-SCHUHMANN2010} and transfer learning~\cite{ruder2019neural} has been shown to increase the recognition rates in such cases~\cite{Giorgi2018}.
Silver standard corpora (SSCs) are annotated corpora that are not annotated by a manual process but rather provide ``suggestive labels'' created by employing distant or weak supervision~\cite{Boland2019}.
In this work, we utilize the Snorkel data programming framework~\cite{ratner2017snorkel} which allows the specification of rules and dictionaries to provide such labels.
The labeling rules are developed based on open knowledge bases and existing literature and generalize to other scientific domains even so they were optimized towards the social science corpus.  
Given those rules, Snorkel trains an unsupervised model for annotation by weighted combination of labeling rules by analyzing the correlations between the matches of the different rules.
Finally, Snorkel provides scores to rank text candidates that were previously extracted from the text by using n-grams with a maximum length which was set to six tokens.
The maximum length was determined from the GSC.
As we optimize the recall, we include all candidates that exceeded the default scoring threshold of .5.
In the following, the distant and weak supervision approach is highlighted in more detail.

\subsubsection{Distant Supervision}
Distant supervision uses external knowledge bases~\cite{mintz2009distant} to retrieve information about candidates of interest.
As Wikidata\footnote{\url{https://www.wikidata.org/}} is recommended for distant supervision~\cite{weichselbraun2019name}, we queried the knowledge graph for software names.
To cover spelling variations and aliases for the different software, we considered various subcategories of software and different software types, and included all aliases from Wikidata's ``Also known as'' attribute in the languages English, German, Spanish and French.
The different languages were included because abbreviations may differ based on the authors' language background even if the articles were written in English.  
We chose these languages as they represent the major languages in Wikipedia.
An overview of Wikidata's variations for SPSS is provided in~\tabref{tab:spss_distant}.
\begin{table}[tb]
    \caption{Aliases for the software `Statistical Package for the Social Sciences'.}
    \label{tab:spss_distant}
    \centering
    \begin{tabularx}{.8\textwidth}{p{3cm}X}
        \toprule
        Wikidata category & Label\\
        \midrule
        label (english) & Statistical Package for the Social Sciences\\
        alias (english) & SPSS\\
        alias (german) & PASW Statistics; PASW\\
        alias (french) & SPSS Inc.; PASW\\
        \bottomrule
    \end{tabularx}
\end{table}
Variations are considered as potential candidates if they do not appear in the regular English dictionary.
Using the English dictionary for exclusion of potential candidates was successfully used by Duck et al.~\cite{duck2013bionerds}.

\newcommand\dl[1]{\protect\dashuline{#1}}
\newcommand\dtl[1]{\protect\dotuline{#1}}
\subsubsection{Weak Supervision with Context Rules}
In addition to the use of external knowledge bases, we implemented labeling functions based on the context of the software usage statements.
To this end, we distinguished between general and exact context rules.
The first examines the context of a candidate for special words or phrases indicating a software mention based on head word rules~\cite{duck2013bionerds}, while the latter implements the set of rules resulting from training an iterative bootstrapping for software identification~\cite{pan2015assessing}.
The general rules employ information about the presence of particular tokens in the context of the candidates, such as `software', `tool' or `package' or the presence of version numbers such as `v0.3', `version 2' or `2.0.12'.
Furthermore, a rule is used that scans for the presence of the developer's name in the context of the candidate.
The identification of the potential candidate's context was done after stop word removal.
Part of Speech tags were employed for selecting from overlapping n-grams.
The exact context rules that determine the context based on a specific pattern are based on the literature~\cite{pan2015assessing}.
Examples for the top two rules are: \verb+use <> software+ and \verb+perform use <>+ where the software position is marked by \verb+<>+.
Example~\ref{ex3} illustrates both the positive application of general context rule (dashed line) and the exact context rule (dotted line).
The exact context rules are applied on the lemmatized context on the training set as in the original rules.
Of the top 10 exact rules, the top 8 were used because the others did not extract any true positives.
\begin{ex}\label{ex3}
We \dtl{used} \ul{SPSS} \dtl{software} \dl{version 23} (\dl{SPSS Inc., Chicago, USA}) for non-image-based statistical analyses and to compare volumes of subcortical structures.
\end{ex}

\subsubsection{SSC Retrieval and Tagging}
Snorkel's generative model was used in its default configuration to generate the suggestive annotations and not further fine-tuned. 
As a final rule the most common false positive n-grams in the training corpus with no true positives were negatively weighted.
To assess the quality of the suggestive labels, the Snorkel generative model was applied on both the training and development corpus and evaluated against the gold standard annotated labels with Snorkel's internal evaluation. 
We trained the Snorkel model on the training set of our corpus and tested if adding further unlabeled data improves the results, since Snorkel is able to learn unsupervised.
However, it was observed that the quality did not improve when adding up to five times the size of the original dataset. 
Overall, the Snorkel model achieved a precision of .33 (.32), a recall of .69 (.64) and an F-score of .45 (.42) on the development (training) set.

The articles for the SSC were obtained by retrieving all articles from PLoS for the keyword ``Social Science'' on 27th of August 2019.
As for the GSC, the M\&M sections were extracted and tagged by Snorkel's model, resulting in a corpus of 51,165 labeled documents.
In total 282,650 suggestive labels were generated for the entire corpus.
\exref{ex4} illustrates a correctly tagged new sample and one partially correct example where a redundant tag was inserted. 
\begin{ex}\label{ex4}
All statistical procedures were performed using \ul{IBM SPSS Statistics} software version 22. 
Task accuracy and response times were analyzed using the \ul{SPSS} software package (\ul{SPSS} v17.0, Chicago, Illinois, USA).
\end{ex}

\section{Extraction of Software Mentions}

\subsection{Model}
For the extraction of software from scientific articles we used a bidirectional Long Short Term Memory Network~\cite{hochreiter1997long} in combination with a Conditional Random Field Classifier~\cite{lafferty2001conditional} (bi-LSTM-CRF) derived from the description by Lample et al.~\cite{lample2016neural}.
This model achieves state of the art performance for Named Entity Recognition.
We used a feature vector consisting of:
\begin{inparaenum}[1.)]
    \item pretrained word-embeddings from scientific publications~\cite{Pyysalo2013DistributionalSR}, and 
    \item bi-LSTM based character-embeddings.
\end{inparaenum} 
Word embeddings capture multi-level semantic similarities between words~\cite{mikolov2013distributed} while character based features allow learning from the orthography of words directly.
The input layer is followed by a bidirectional LSTM layer to consider the surrounding context of software mentions, a fully connected layer for classification, and a final CRF layer for the estimation of the most likely tagging sequence~\cite{lample2016neural}.
The model's hyper-parameters are summarized in \tabref{tab:model_config}.
\begin{table}[bt]
    \caption{Summary of the extraction model hyper-parameter settings.}
    \label{tab:model_config}
    \centering
    \begin{tabular}{ll}
        \toprule
        Hyper-parameter & Setting  \\  
        \midrule
        Word Embedding Size & 200 (pre-trained, not trainable)\\ 
        Character Embedding Size & 50 (trainable)\\
        Character LSTM Size & 25 \\
        Main LSTM Size & 100 \\
        Number of Labels & 3 (`O', `B-software', `I-software') \\
        \bottomrule
    \end{tabular}
\end{table}

\subsection{Training}
As discussed above, training of the model was based on two different corpora, an SSC and a GSC.
Sequential transfer learning~\cite{ruder2019neural} was employed to transfer information learned from the suggestive labels of the SSC to the GSC with the high quality labels to cope with the small amount of training data.
For SSC training we selected all positive samples per epoch and used one negative sample per positive sample. 
We used different negative samples for each epoch until all were seen once. 
The number of pre-training epochs was optimized by training with up to 25 consecutive epochs on the SSC, after each of which we tested the performance to be expected by applying a standard, optimized re-training routine on the GSC training set. 
This procedure was selected due to the high computational requirements.
For the SSC training, 2 epochs were found to provide the best basis.
To optimize the SSC training, different learning and drop-out rates were considered.
The best performing model was then selected for further optimization of the GSC training.
All optimizations and evaluations were done with \textit{rmsprop} to perform stochastic gradient descent. 

During GSC training optimization, the following hyper-parameters were systematically considered:
\begin{inparaenum}[1.)]
    \item \textit{drop-out rate} in range of .3--.6,
    \item \textit{learning rate} in the range of .0001--.003,
    \item \textit{learning rate decay}, and 
    \item \textit{sample weighting} adjusts the loss function label specific in a range of 0--.2 to increase the weight of positive samples.
\end{inparaenum}
For GSC training we stop the training after 22 epochs.
The final hyper-parameters for SSC and GSC training are provided in \tabref{tab:training_config}.
\begin{table}[tb]
    \caption{Hyper-parameter settings for the training process.}
    \label{tab:training_config}
    \centering
    \begin{tabularx}{.8\textwidth}{p{3cm}XX}
        \toprule
         Hyper-parameter & SSC & GSC \\
        \midrule
        Learning Rate & .002  & .0015\\
        Learning Decay & .0001 (linear) & .0007 (exponential)\\
        Dropout Rate & .5 & .4 \\
        Sample Weight & .1 & .1\\
        Epochs & 2 & 22 \\
        \bottomrule
    \end{tabularx}
\end{table}

\subsection{Evaluation and Extraction}
To determine the expected quality of the final software mention identification, the selected model's performance was evaluated in precision, recall and F-score on both, the development and the test set.
We consider both precision and recall as highly important for the intended applications of the model. 
Precision allows, for instance, to give accurate impact measures while a high recall is beneficial for discovering rare, domain specific software. 
For testing on the development set the final model was trained on the training set alone while for testing on the test set the model was trained on both the training and development sets. 
This approach enables the estimation of the influence additional training data has. 
There are four relevant evaluation methods:
\begin{inparaenum}[1.)]
    \item B-software: a match of the first token in a software name,
    \item I-software: a match of all other words in a software name except for the first, 
    \item partial match: an overlap of the estimated and the true software name, and
    \item exact match: the exact identification of the entire software name. 
\end{inparaenum}
To also assess the effect of the manually annotated data and the transfer learning, the following evaluations were performed:
\begin{inparaenum}[1.)]
    \item SSC: the model resulting from SSC training only,
    \item GSC: the model resulting from GSC training only, and 
    \item SSC$\rightarrow$GSC: the final model resulting from transfer learning based on SSC and GSC training.
\end{inparaenum}
The performance scores are summarized in \tabref{tab:results}.
For comparison we also tested a CRF based on a set of standard features\footnote{\url{https://sklearn-crfsuite.readthedocs.io}} on the GSC which achieved an F-Score of .41 (.36) with a precision of .30 (.24) and a recall .66 (.70) and  on test (devel) set for exact recognition. 
\input{results}

As we found an increase in the recognition performance from increasing the amount of training data for all evaluation metrics, the model used for information extraction was trained on the full GSC.
It was then applied to all 51,165 M\&M section from PLoS to identify software usage statements, resulting in 133,651 software names, 25,900 of which are unique.
This seems plausible, as it reflects a similar frequency of mentions per article (2.6) as in the GSC (2.9).    

\subsection{Entity Disambiguation and Additional Information}
Software names in scientific literature contain many variations in spelling and level of detail.
This ranges from including the manufacturer's name to using version information and different interpretations of software name abbreviations.
\tabref{tab:spss_variations} gives an overview of the most common spelling variations of the statistical software SPSS.
In total, 179 different spellings for SPSS were identified, most of which were not in the GSC.
This means the model is able to generalize to previously unknown software names. 
\begin{table}[tb]
    \centering
    \caption{Overview of the most frequent spelling variations for SPSS.}
    \label{tab:spss_variations}
    \begin{tabularx}{.72\textwidth}{p{7cm}p{2cm}}
        \toprule
        Software mention & Frequency\\
        \midrule
        SPSS & 7068\\
        Statistical Package for the Social Sciences & 944\\
        IBM SPSS Statistics & 875\\
        Statistical Package for Social Sciences & 784\\
        IBM SPSS & 480 \\
        \bottomrule
    \end{tabularx}
\end{table}

To disambiguate the different spelling variations, a three part entity linking was employed to reflect software mention specific variations:
\begin{inparaenum}[1.)]
    \item analysis of simple spelling variations based on mentions,
    \item abbreviation based linking, and
    \item exploitation of information from the DBpedia knowledge graph.
\end{inparaenum}
Software mentions were normalized by case folding and removal of special characters such as numbers and Greek letters.
Furthermore, syllables such as ``pro'' were removed from the end of software names and the remaining words were stemmed to match common variations such as `statistical' and `statistic'.
The result of this step was a transformed version of the software mention.
To match the different interpretations of abbreviations, stop-words were removed and abbreviations were created from the first letters of the remaining words.
The transformed name and the abbreviations were then used to cluster the first software names in the hierarchical linking pipeline.

For further disambiguation all software from DBpedia including \textit{label}, \textit{name}, \textit{wikiPageRedirects} and \textit{wikiPageDisambiguates} was retrieved and used to link the software names and select unique names as previously suggested~\cite{weichselbraun2019name}. 
Here, programming languages were included and software of type ``video game'' excluded.
As for the Wikidata query, the languages English, German, French and Spanish were considered.
If available the developer of a software was also included.
The actual linking was performed as follows:
\begin{compactenum}[1.)]
    \item labels from DBpedia were used to further group spelling variations,
    \item aliases from all languages were used, and
    \item if neither of the previous provided a match, a combination of label and developer was employed.
\end{compactenum} 
The representative name used in the final data model was then created from the DBpedia label of the software, if exists, or the most frequent matched name, otherwise.
DBpedia entries were retrieved for 66,899 (1160 unique) of the software names.
As a result of the entity disambiguation, the set of unique software names was reduced from 25,900 to 20,227.

If available, the following additional information was collected in a manual process for the most frequent software:
\begin{inparaenum}[1.)]
    \item the corresponding identifier in the Software ontology, Wikidata, and Wikipedia,
    \item the URL of the software,
    \item the manufacturer of the software,
    \item whether the software is freely available,
    \item whether the source code of the software is freely available, and
    \item the license that was used for the publication of the source code.
\end{inparaenum}
Additional information was retrieved for 133 software products covering 67,477 software mentions, representing half of all software mentions.

\section{SoftwareKG: Data Model and Lifting to RDF}

SoftwareKG was generated from the extracted and additionally collected information.
In particular, it contains metadata about the publications, the authors and the software used in the publications.
The graph contains 3,998,194 triples, 1,013,216 resources which are represented in 5 distinct types and 25 distinct properties.
The graph holds 51,165 \textit{Publication} resources, 20,227 \textit{Software} resources, 334,944 \textit{Author} resources, and 473,229 \textit{Organization} resources.
In total, the graph contains 133,651 software mentions.
The most frequent software mentioned in the papers are \textit{SPSS} (11,145 mentions), \textit{R} (11,102 mentions), and \textit{STATA} (5,783 mentions).
Of the included software, 75 software applications are available for free, while 58 are not free.
For the remaining software, this information is missing.
Similarly, the source code is available of 52 software applications.

The SoftwareKG data model exploits terms from established vocabularies, mostly from schema.org\footnote{\url{https://schema.org/}}. 
For designing the data model, established vocabularies have been reused as it is seen as best practice~\cite{heath2011linked}.
The model can easily be extended by further properties, e.g. by terms of the CodeMeta Project\footnote{\url{https://codemeta.github.io/}} for describing software metadata.
For some very specific properties, we had to define our own properties which are denoted in the model by the namespace \texttt{skg}.

\begin{figure}[t]
    \centering
    \includegraphics[width=\textwidth]{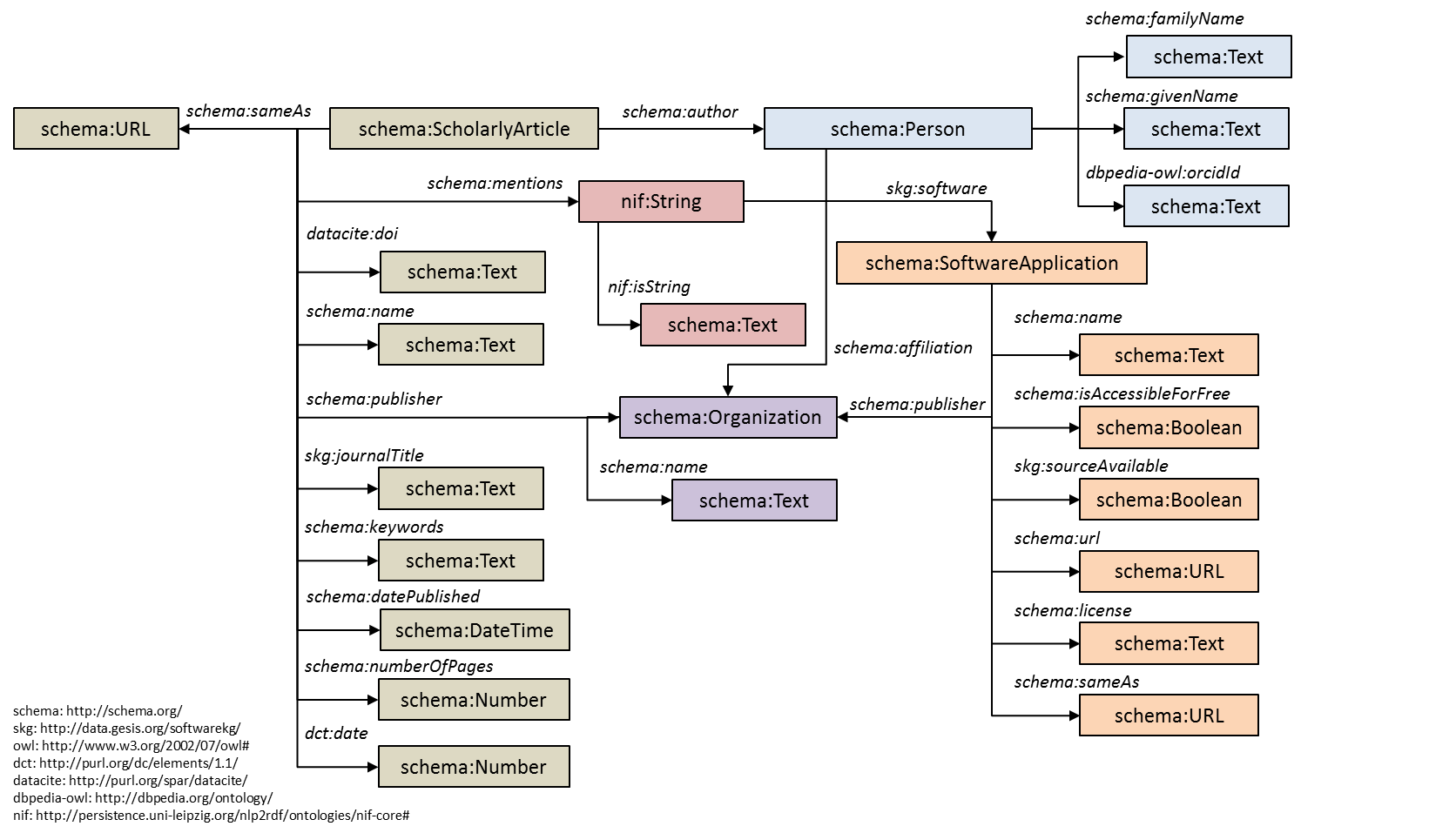}
    \caption{Illustration of the Data Model}
    \label{fig:data_model}
\end{figure}

The core elements of the model are \textit{Software}, \textit{Publication},  \textit{Author}, and \textit{Organization} as shown in \figref{fig:data_model}. 
The class \textit{Software} represents a software application. 
It gathers properties for representing the name of the software application, the publisher, the homepage, the license under which the software has been published, whether it is available for free, and whether the source code is available or not.
The property \texttt{schema:sameAs} gathers links to Wikipedia, Wikidata, the Software Ontology, and to DBpedia.
The class \textit{Publication} represents a scientific article. 
Here, we use properties to represent title, author(s), DOI, publisher, the publication date, and other metadata.
We link to the same publication in the Microsoft Academic Graph by using \texttt{schema:sameAs}.
The property \texttt{schema:mentions} captures the detected software mentions in a publication.
Each mention is from the type \texttt{nif:String} which connects to the precise string in the paper (\texttt{nif:isString}) and to the meant \textit{Software} (\texttt{skg:software}).
The class \textit{Author} represents each of the authors of a publication with his/her name, affiliation, and Orcid ID.
Eventually, to the class \textit{Organization} it is linked to by authors (as their affiliation), publications (as their publisher), and software (also as publisher).

All extracted and linked information is generated in JSON-LD following the data model. 
SoftwareKG is published under the Creative Commons BY 4.0 license.
The KG can be accessed from a Virtuoso triple store with a SPARQL endpoint\footnote{\url{https://data.gesis.org/softwarekg/sparql}} and is downloadable as a dump\cite{Schindler_softwarekg_2020}.
It can also be accessed through its official website\footnote{\url{https://data.gesis.org/softwarekg/site/}} which also contains statistics and a set of SPARQL queries.

\section{Use Cases and Exploitation}
As initially stated, knowledge about the software employed in scientific investigations is important for several reasons.
The previous sections describe how SoftwareKG, a knowledge graph about software mentions in scientific publication was created.
This section illustrates the usage of SoftwareKG and how to leverage information from other knowledge graphs to perform analyses about software in science. 
The queries and the plots illustrating their results are available from the accompanying website

\begin{figure}[tb]
    \centering
    \includegraphics[width=\textwidth]{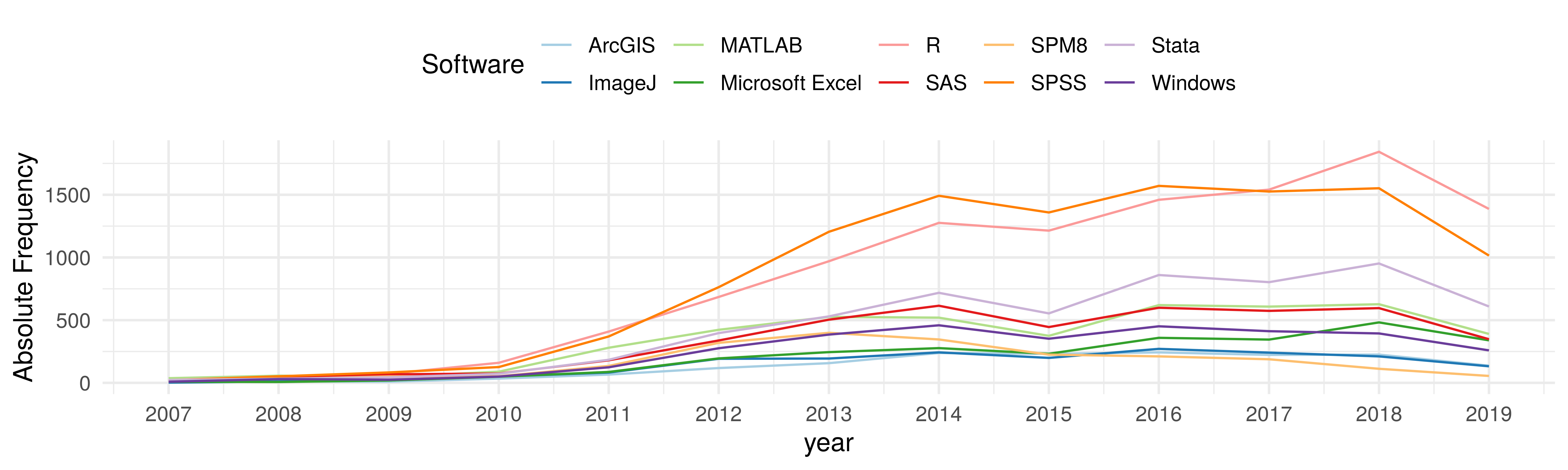}
    \caption{Absolute amount of the 10 most common software per year.}
    \label{fig:counts}
\end{figure}

The frequency of software mentions in scientific articles allow to assess the impact of individual software to science.
Moreover, it allows to attribute the developers of such software.
\figref{fig:counts} illustrates the frequencies of the 10 most common software per year in absolute numbers with respect to our corpus.
The data was obtained by using the query in Listing~\ref{lst:sparql}.
It can be seen that both, SPSS and R are predominantly used in the social sciences, reflecting their usage for statistical analyses.
In general, statistical analysis software is the most frequently used type of software.

\begin{figure}[tb]
    \centering
    \includegraphics[width=\textwidth]{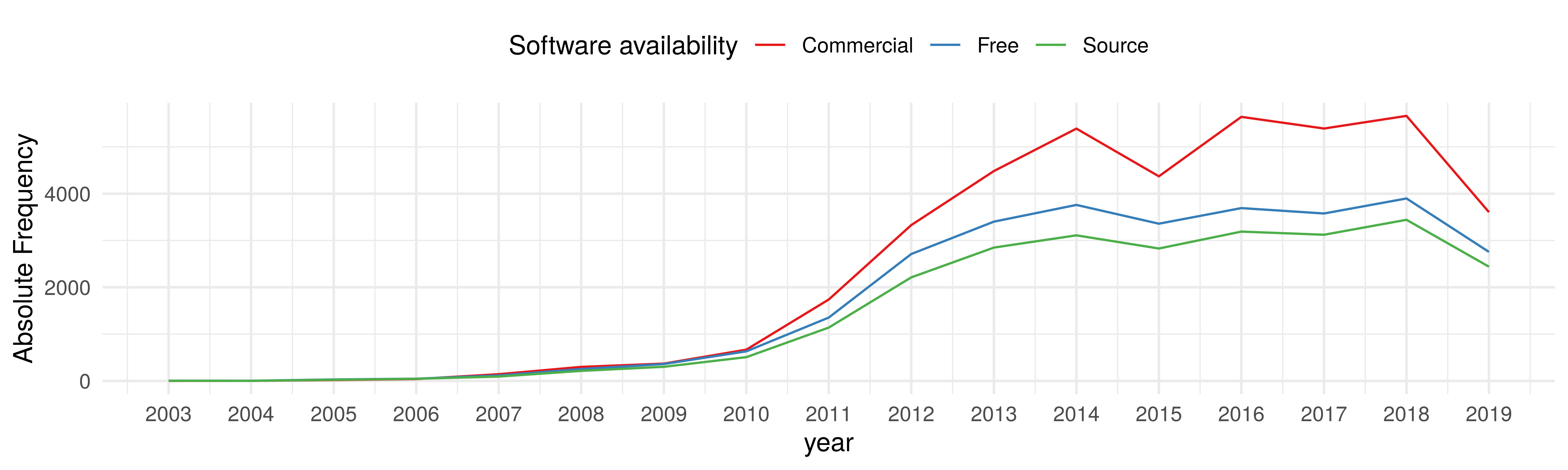}
    \caption{Absolute frequency of the usages of commercial, free, and open source software.}
    \label{fig:open_free}  
\end{figure}

The availability of software used for the original analyses of scientific investigations plays a central role in its reproducibility.
Moreover, the usage of open source software allows researchers to inspect the source code, reducing uncertainty about the reliability of the scientific analyses and the results~\cite{Russo2016}.
\figref{fig:open_free} illustrates the usage of free and open source software over time.

\begin{figure}[tb]
    \centering
    \includegraphics[width=\textwidth]{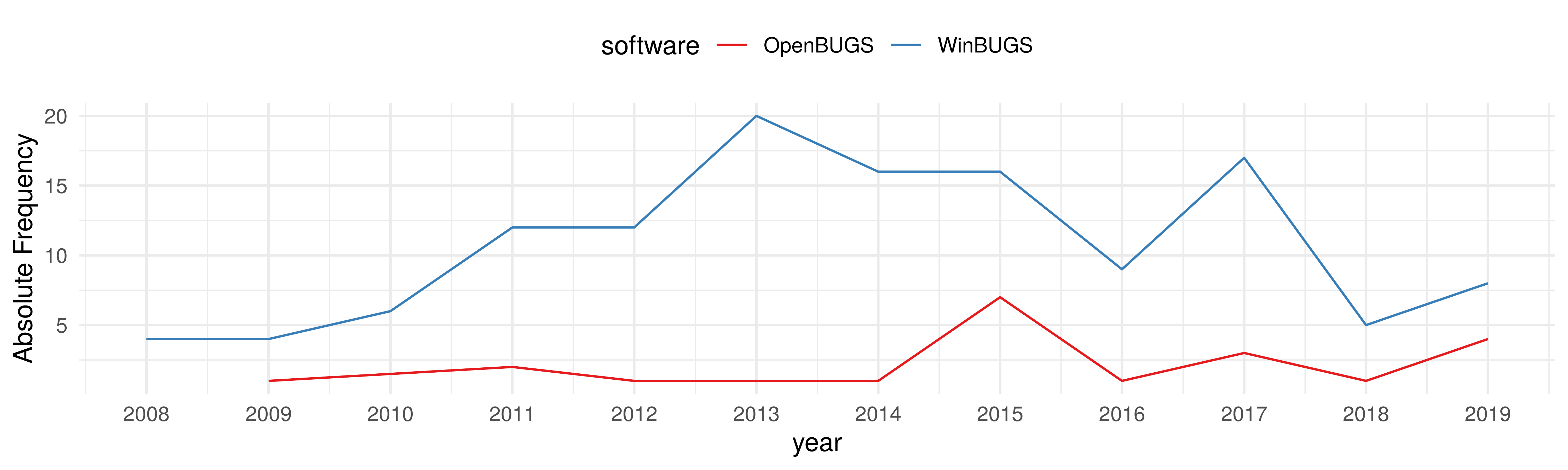}
    \caption{Absolute frequency of the usages of OpenBUGS and WinBUGS.}
    \label{fig:BUGS}
\end{figure}

When the development of a software is discontinued, it is often replaced by another software.
This is, for instance, the case with the free software WinBUGS, with the latest release from 2007 and the open source software OpenBUGS, where the development started in 2005.
This relation can easily be retrieved via Wikidata using the \textit{replaced-by}(P1366) relation.
Through linking SoftwareKG with WikiData we combined information about the replacement of software with knowledge about its usage in scientific articles which enables statements about when and how such transitions arrive in science. 
The query linking SoftwareKG and WikiData is available on the SoftwareKG website and with the source code~\cite{Schindler_softwarekg_2020}.
\figref{fig:BUGS} illustrates the frequencies of both the free software WinBUGS and the open source software OpenBUGS per year.
From the plot, it can be seen that while the development of WinBUGS was discontinued more than 10 years ago, the successor did not replace it yet.

\lstset{language=SQL,morekeywords={PREFIX,java,rdf,rdfs,url}}
\begin{lstlisting}[float,captionpos=b, caption={SPARQL query to retrieve frequency of software mention per year}, label={lst:sparql}, basicstyle=\ttfamily,frame=single]
PREFIX schema: <http://schema.org/>
SELECT ?n ?y (count(?n) as ?count) WHERE {
    ?s rdf:type schema:SoftwareApplication .
    ?s <http://schema.org/name> ?n .
    ?m <http://data.gesis.org/softwarekg/software> ?s .
    ?p <http://schema.org/mentions> ?m .
    ?p <http://purl.org/dc/elements/1.1/date> ?y .
}
GROUP BY ?n ?y
HAVING (count(?n) > 1)
ORDER by DESC(?count)
\end{lstlisting}

\section{Related Work}\label{sec:sota}
Identifying software mentions in scientific publications is an open research problem and can serve several different purposes: 
\begin{inparaenum}[1.)]
\item mapping of available software, 
\item measuring the impact of software, and 
\item analyzing how software is used, shared and cited in science. 
\end{inparaenum}
A more detailed overview of the problem and the reasons can be found in~\cite{Krueger2019}.
Besides manual extraction, as for example done by Howison et al.~\cite{howison2016software} to investigate software citations and their completeness, automated extraction enables the analysis of software usage in a larger context. 
Greuel and Sperber~\cite{greuel2014swmath} create a mapping of mathematical software through automated filtering of potential mentions and manual review.
Duck et al.~\cite{duck2013bionerds} use a fully automated, rule-based scoring system configured by hand and based on a dictionary of known software and databases which was later improved by applying machine learning onto the rule set to achieve  .67 F-score in the domain of Bioinformatics~\cite{duck2016survey}. 
Pan et al.~\cite{pan2015assessing} use iterative bootstrapping to learn common mentioning patterns and software names starting from an initial set of seed software entities achieving .58 F-score and use the extraction results to provide impact measures. 
While those investigations all deal with the extraction of software they are not concerned with entity linking and can therefore only argue about distinct software mentions. 

Automatically generating knowledge graphs from information about scientific publications allows the analysis of scientific workflows and enable large scale meta-analyses. 
The Open Academic Graph\footnote{\url{https://www.openacademic.ai/oag/}}, for instance, captures metadata of scientific publications, while the Scholarlydata project~\cite{nuzzolese2016conference} captures linked data about scientific conferences and workshops.
Jaradeh et al.~\cite{jaradeh2019open} create the Open Research Knowledge Graph which captures semantic information from scientific publications. 
They apply deep learning methods to capture information about process, method, material and data in publications and use DBpedia for entity linking. 
However, they do not provide a quantitative evaluation of their text mining approach. 
Recupero et al.~\cite{recupero2019mining} employ different existing classifiers to extract knowledge from scientific articles, perform disambiguation between extracted targets and create a knowledge graph based on the gathered information. 
Luan et al.~\cite{luan2018multitask} create a knowledge graph from tasks, methods, metrics, materials, and other entities and their relations from scientific publications.

The work presented here is the first that creates a knowledge graph particularly tailored for analyses of software in science based on recognition performance that outperforms previous approaches and the first to include entity disambiguation for software mentions.

\section{Limitations and Potential Sources of Uncertainty}\label{sec:limits}
Most of the results presented in this work are based on methods of machine learning and automatic analyses which mainly rely on the quality of the provided labeled corpus.
For the entire pipeline we identified several sources of uncertainty, which might accumulate and may result in a bias for further analyses:

\begin{itemize}
    
\item The corpus was retrieved from PLoS by using the keyword ``Social Science'' potentially resulting in the following two issues: The employed keyword did not only result in articles from the social sciences, but also from the bio medicine and related research domains. 
On the other hand this aspect facilitates the transfer of the model to the domain of life sciences.
Additionally, the PLoS corpus itself contains a bias towards the open access general purpose journal, which might not reflect the preferred publication target of researchers from the target domain.
    
\item While the GSC annotation is of high quality in terms of inter-rater reliability the annotation task proved to be a complex task for the annotators, for instance when differentiating between algorithms and software with the same name. 
The absence of version information makes this decision even harder.
    
\item The SSC is constructed with suggestive labels, with false positives that may be carried over to the final model.
Indeed, we found examples, such as \textit{Section} and \textit{ELISA}, which both are software names, but also commonly appear in scientific publications without a connection to the software. 

\item The employed model achieves a high recognition performance.
As suggested by the improvement of the test set evaluation, there is still potential to increase the performance by using a larger GSC.

\item Reliably estimating the error of the entity disambiguation is difficult due to the absence of a ground truth. 
However, the method benefits from just working on extracted names which strongly restricts the chance of errors. 
We found some cases in which a linking to DBpedia was not possible because of multiple matches with DBpedia entries for which we could not automatically determine which match is the correct one. 
\end{itemize}

In summary, SoftwareKG provides a reasonable quality in terms of knowledge identification, but automatic analyses should carried out carefully.

\section{Conclusion and Future Work}
In this work we introduce SoftwareKG and a method for creating a large scale knowledge graph capturing information about software usage in the social science.
It was generated by employing a bi-LSTM for the automatic identification of software mentions in the plain text of publications. 
The proposed method achieves a high recognition score of .82 F-score in terms of exact match, which is a strong improvement over the state of the art.
By using transfer learning based on data programming with distant and weak supervision rules, the performance could be significantly improved.
The proposed approach is the first to integrate entity linking for the disambiguation of software names and the first to construct a knowledge graph to facilitate reasoning by allowing running queries against the constructed graph. 
Additionally, the available information in our graph was enhanced by manual annotation to support further analyses regarding free and open source software where otherwise no analysis would be possible. 
To create a large scale basis for reasoning and illustrate how such a basis can be constructed we applied our method to construct a knowledge graph over all articles published by PLoS tagged with ``Social Science''.
Finally, we employed SoftwareKG to illustrate potential analyses about software usage in science.

Future work includes the automatic collection of additional information about the software such as the version or the source code repositories which implies an extension of the data model, e.g. by including properties of CodeMeta.
This enables the identification of the particular implementation by employing software preservation services such as SoftwareHeritage\footnote{\url{https://www.softwareheritage.org/}}.

\section*{Acknowledgements}
This work was partially carried out at GESIS - Leibniz Institute for the Social Sciences and was financially supported by the GESIS research grant GG-2019-015 and by the Deutsche Forschungsgemeinschaft (DFG, German Research Foundation) - SFB 1270/1 - 299150580.

\bibliographystyle{splncs04}
\bibliography{references}

\end{document}

%% file: results.tex
\begin{table}[tb]
    \centering
    \caption{Overview of the recognition results of the software identification.}
    \label{tab:results}
    \begin{tabularx}{\textwidth}{XXp{1cm}p{1cm}p{1cm}p{1cm}p{1cm}p{1cm}}
      \toprule
    Evaluation & Training & \multicolumn{2}{c}{Precision}  & \multicolumn{2}{c}{Recall} & \multicolumn{2}{c}{F-score} \\ 
    && test & dev & test & dev & test & dev\\
      \midrule
       & SSC & .21 & (.21) & .72 & (.70) & .32 & (.32) \\
      B-software & GSC & .83 & (.75) & .74 & (.68) & .78 & (.71) \\ 
       & SSC$\rightarrow$GSC & \textbf{.86} & (.83) & \textbf{.85} & (.78) & \textbf{.86} & (.80) \\ 
      \midrule
       & SSC & .36 & (.31) & .68 & (.35) & .47 & (.33) \\ 
      I-software & GSC & \textbf{.86} & (.75) & .66 & (.47) & .75 & (.58) \\ 
       & SSC$\rightarrow$GSC & .76 & (.77) & \textbf{.82} & (.61) & \textbf{.79} & (.68) \\ 
      \midrule
      & SSC & .21 & (.22) & .72 & (.74) & .32 & (.34) \\ 
      partial match & GSC & .85 & (.75) & .76 & (.69) & .80 & (.72) \\ 
       & SSC$\rightarrow$GSC & \textbf{.87} & (.84) & \textbf{.85} & (.80) & \textbf{.86} & (.82) \\ 
      \midrule
       & SSC & .20 & (.20) & .68 & (.64) & .30 & (.30) \\ 
      exact match & GSC & .80 & (.72) & .72 & (.66) & .76 & (.69) \\ 
       & SSC$\rightarrow$GSC & \textbf{.83} & (.81) & \textbf{.82} & (.78) & \textbf{.82} & (.79) \\ 
       \bottomrule
    \end{tabularx}
    \end{table}